\documentclass[aps,preprint,showpacs,floatfix]{revtex4-1}
\usepackage{bm}
\usepackage{amsmath}
\usepackage{epsfig}
\usepackage{simplewick}
\usepackage{rotating}
\usepackage{fancyhdr}
\usepackage[T1]{fontenc}
\usepackage{times}

\def\lazz{\mathrel{\mathchoice {\vcenter{\offinterlineskip\halign{\hfil
$\displaystyle##$\hfil\cr<\cr\sim\cr}}}
{\vcenter{\offinterlineskip\halign{\hfil$\textstyle##$\hfil\cr<\cr\sim\cr}}}
{\vcenter{\offinterlineskip\halign{
\hfil$\scriptstyle##$\hfil\cr<\cr\sim\cr}}}
{\vcenter{\offinterlineskip\halign{\hfil$\scriptscriptstyle##
$\hfil\cr<\cr\sim\cr}}}}}
\def\gazz{\mathrel{\mathchoice {\vcenter{\offinterlineskip\halign{\hfil
$\displaystyle##$\hfil\cr>\cr\sim\cr}}}
{\vcenter{\offinterlineskip\halign{\hfil$\textstyle##$\hfil\cr>\cr\sim\cr}}}
{\vcenter{\offinterlineskip\halign{
\hfil$\scriptstyle##$\hfil\cr>\cr\sim\cr}}}
{\vcenter{\offinterlineskip\halign{\hfil$\scriptscriptstyle##
$\hfil\cr>\cr\sim\cr}}}}}		
\def\pr{\prime}
\def\be{\begin{equation}}
\def\lan{\left\langle}
\def\ran{\right\rangle}
\def\ee{\end{equation}}
\def\barr{\begin{array}}
\def\earr{\end{array}}

\def\l{\left}
\def\r{\right}
\def\dis{\displaystyle}
\def\ed{\end{document}}

\def\co{{\cal O}}
\def\ch{{\cal H}}
\def\can{{\cal N}}

\def\cg{{\cal G}}

\def\cn{{\bf n}}

\def\spin{\frac{1}{2}}
\def\tmp{\widetilde{m_p}}
\def\tmn{\widetilde{m_n}}
\def\dg{\dagger}
\def\wm{{\widetilde {m}}}
\def\cf{{\cal F}}
\def\wtM{{\widetilde {M}}}

\def\bcon{\bcontraction}

\begin{document}

\title{Spectral distribution method for neutrinoless double-beta decay
nuclear  transition matrix elements: Binary correlation results}

\author{Manan Vyas$^1$ and V.K.B. Kota$^{1,2,}$
\footnote{ Corresponding author, phone:
91-79-26314464, Fax: +91-79-26314460 \\ {\it E-mail address:} 
vkbkota@prl.res.in (V.K.B. Kota)}}

\affiliation{$^1$Physical Research Laboratory, Ahmedabad 380 009, India \\
$^2$Department of Physics, Laurentian University, Sudbury, Ontario, Canada
P3E 2C6}

\begin{abstract}

Neutrinoless double-beta decay nuclear transition matrix elements are
generated by an effective two-body transition operator and it consists of 
Gamow-Teller like and Fermi like (also tensor) operators. Spectral
distribution method for the corresponding transition strengths (squares of
the transition matrix elements) involves convolution of the transition
strength density generated by the non-interacting particle part of the
Hamiltonian with a spreading function generated by the two-body part of the
Hamiltonian. Extending the binary correlation theory for spinless embedded
$k$-body ensembles to ensembles with proton-neutron degrees of freedom, we
establish that the spreading function is a bivariate Gaussian for transition
operators $\co(k_\co)$ that change $k_\co$ number of neutrons to $k_\co$
number of protons. Towards this end, we have derived the formulas for the
fourth-order cumulants of the spreading function and calculated their values
for some heavy nuclei; they are found to vary from $\sim -0.4$ to
$-0.1$. Also for nuclei from $^{76}$Ge to $^{238}$U, the bivariate
correlation coefficient is found to vary from $\sim 0.6 - 0.8$ and these
values can be used as a starting point for calculating nuclear transition
matrix elements using the spectral distribution method.

\end{abstract}


\maketitle

\section{Introduction}

Double-$\beta$ decay (DBD) is an extremely rare weak-interaction process in
which two identical nucleons inside the nucleus undergo decay with or
without emission of neutrinos. Theoretically, the two neutrino double beta
decay ($2\nu\beta^-\beta^-$) process was first predicted by Mayer
\cite{Mey-35} following the suggestion of Wigner. This process has been
observed in more than 10 nuclei and best adopted values for the half-lives
were tabulated recently by Barabash \cite{Ba-10}. In 1937, following the
suggestion of E. Majorana \cite{Ma-37},  Racah \cite{Ra-37} pointed out the
possibility of neutrinoless double-$\beta$ decay (NDBD or
$0\nu\beta^-\beta^-$). Furry \cite{Fu-39} in 1939, for the first time
calculated NDBD  half-lives. Fundamental significance of NDBD is that its
experimental confirmation will tell us about lepton number violation in
nature and that neutrino is a Majorana particle. More importantly, NDBD
gives a value or a bound on neutrino mass \cite{En-08} provided the 
half-lives are known experimentally and the corresponding nuclear transition
matrix elements (NTME) are obtained using a reliable nuclear model. So far
only Klapdor et al \cite{Kl-04} claimed to have evidence (at a confidence
level of $4.2\sigma$) for $0\nu\beta^-\beta^-$ in $^{76}$Ge. At present
large number of NDBD experiments are being carried out and many others are
in development stage in various laboratories around the  world. The nuclei
being considered are $^{48}$Ca, $^{76}$Ge, $^{82}$Se, $^{100}$Mo,
$^{116}$Cd, $^{130}$Te,  $^{136}$Xe, $^{150}$Nd and so on \cite{En-08}.
Following this, several nuclear models are employed for calculating the NTME
for various candidate nuclei mentioned above.  Some of the models used for
NDBD studies are shell model using recent state-of-the-art large scale
calculations \cite{Cau-08, Men-09}, quasi-particle random-phase
approximation with various extensions \cite{Su-98, Bob-00}, interacting
boson model \cite{Bar-09}, pseudo-SU(3) model \cite{Hi-02}, projected
Hartree-Fock-Bogoliubov method including pairing plus quadrupole-quadrupole
interaction \cite{Cha-08, Rath-09}, generating coordinate method with
particle number and angular momentum projection \cite{Rod-10}. 

Statistical spectral distribution theory \cite{KH-10} gives a method for
calculating transition strengths (squares of transition matrix elements)
generated by a transition operator. This theory starts with shell model
spectroscopic spaces and the same shell model inputs (single particle
energies and effective two-body interactions). Here one constructs smoothed
forms (spectral distributions) for various observables ignoring the
fluctuations and this is based on random matrix representation of
Hamiltonians (also other operators), unitary decompositions of the operators
and quantum chaos. Spectral distribution theory has been applied in the
past, with various approximations, to a variety of problems in nuclear
structure and they include (i) bound on time-reversal non-invariant part of
the nucleon-nucleon interaction \cite{FKPT}, (ii) single particle transfer
\cite{Po-91}, (iii) $\beta$-decay rates for pre-supernovae evolution
\cite{Ko-95,Go-11}, (iv) giant dipole widths \cite{Ma-98}, and (v) parity
breaking matrix elements in compound nuclei \cite{To-01}. Our purpose in
this paper is to describe spectral distribution method for the NTME for NDBD
and establish that the spreading function that enters in the theory is close
to a bivariate Gaussian. With this result, it is possible to apply in future
spectral distribution method to NDBD. Now we will give a preview.

In Sec. II, we describe briefly the structure of the  $0\nu\beta^-\beta^-$
transition operator which is a two-body operator. Section III describes the
spectral distribution method for NDBD. Section IV gives the binary
correlation theory for traces of operators over two-orbit (proton and
neutron) configurations. Using these, in Sec. V, derived are the results for
the fourth-order bivariate cumulants and the bivariate correlation
coefficient for the spreading transition strength density function
appropriate for NDBD. Numerical results are presented for the fourth-order
cumulants to demonstrate that the transition strength density is close to a
bivariate Gaussian. In addition, for several heavy nuclei, the values for
the bivariate correlation coefficient are presented. Finally, Sec. VI gives
conclusions.

\section{0$\nu$ double-beta decay transition operator}

Half-life for 0$\nu$ double-beta decay, for the 0$^+_i$ ground state
(gs) of a initial even-even nucleus decay to the 0$^+_f$ gs of the final
even-even nucleus is given by \cite{El-02}
\be
\l[ T_{1/2}^{0\nu}(0^+_i \to 0^+_f) \r]^{-1} =  G^{0\nu}
\l| M^{0\nu} (0^+)\r|^2 \l(\dis\frac{\lan m_\nu \ran}{m_e}\r)^2 \;,
\label{eq.dbd1}
\ee
where $\lan m_\nu \ran$ is the effective neutrino mass (a combination of
neutrino mass eigenvalues and it also involves neutrino mixing matrix).  The
$G^{0\nu}$ is phase space integral  (kinematical factor) dependent on
charge, mass and available energy for NDBD process; tabulations for
$G^{0\nu}$ are available \cite{Bo-92,Doi-93}. The $M^{0\nu}$ is the NTME 
generated by the NDBD transition operator and it is a sum of a Gamow-Teller 
like ($M_{GT}$), Fermi like ($M_F$) and tensor ($M_T$) two-body operators. 
As it is well known that the tensor part contributes only up to 10\% of the 
matrix elements \cite{Ba-09,Ho-10},  we will neglect the tensor part. 
Then we have, from the closure approximation which is well justified for 
NDBD \cite{El-02},
\be
\barr{rcl}
M^{0\nu} (0^+) & = & M^{0\nu}_{GT} (0^+) - \dis\frac{g_V^2}{g_A^2} 
M^{0\nu}_{F} (0^+) = \lan 0^+_f \mid\mid \co(2:0\nu) \mid\mid 0^+_i \ran \;,
\\ \\
\co(2:0\nu) & = & \dis\sum_{a,b} \ch(r_{ab}, \overline{E}) 
\tau_a^+ \tau_b^+ \l( \sigma_a \cdot \sigma_b- \dis\frac{g_V^2}{g_A^2} \r)
\;.
\earr \label{eq.dbd2}
\ee
As seen from Eq. (\ref{eq.dbd2}),  NDBD half-lives are generated by the
two-body transition operator $\co(2:0\nu)$; note that $a,b$ label 
nucleons. The $g_A$ and $g_V$ are the weak axial-vector and vector coupling
constants.  The $\ch(r_{ab}, \overline{E})$ in Eq. (\ref{eq.dbd2}) is
called `neutrino potential'. Here $\overline{E}$ is the average energy of
the virtual intermediate states used in the closure approximation. The form
given by Eq. (\ref{eq.dbd2}) is justified {\it only if the exchange of the
light Majorana neutrino is indeed the mechanism responsible for the NDBD}.
The neutrino potential, defining completely the two-body NDBD transition
operator $\co(2:0\nu)$  is, to a good approximation, 
given by \cite{Ba-09,Ho-10,En-88,To-91,Si-09},
\be
\ch(r_{ab}, \overline{E}) = \dis\frac{R}{r_{ab}}\Phi(r_{ab},
\overline{E})\;.
\label{eq.dbd3a}
\ee 
Here, $R$ in fm units is the nuclear radius and similarly  $r_{ab}$ is in
fm units. The function $\Phi$ is given by \cite{Ba-09,Ho-10,Si-09}, 
\be
\Phi(r_{ab},\overline{E})=
\dis\frac{2}{\pi }\left[ \sin\l(\frac{\overline{E}\,r_{ab}}
{\hbar c}\r)f_{1}\l(\frac{\overline{E}\,r_{ab}}
{\hbar c}\r)-\cos\l(\frac{\overline{E}\;r_{ab}}
{\hbar c}\r)f_{2}\l(\frac{\overline{E}\,r_{ab}}
{\hbar c}\r)\right]\;.
\label{eq.dbd3aa}
\ee 
In Eq. (\ref{eq.dbd3aa}), $f_{1}(x)=-\int_{x}^{\infty }t^{-1}\cos t \,dt = 
Ci(x) = \gamma + \ln \, x + \int_{0}^{x }t^{-1}(\cos t -1) \, dt$ and $f_{2}
(x) = -\int_{x}^{\infty }t^{-1}\sin t \,dt = Si(x) -  \frac {\pi}{2}$;
$Si(x)$ and $Ci(x)$ are the sine and cosine integrals. It is useful to
mention that $ \Phi(r_{ab},\overline{E}) \sim  \exp ({-\frac{3}{2}
\frac{\overline{E}}{\hbar c} r_{ab}})$. Note that $\hbar c =197.327$ MeV
fm.  The effects of short-range correlations in the wavefunctions are
usually taken into  account by multiplying the wavefunction by the Jastrow
function $[1 - \gamma_3e^{-\gamma_1 r_{ab}^2}  ( 1 - \gamma_2 r_{ab}^2 )]$
\cite{Ho-10,En-88,Si-09}. Now keeping the wavefunctions unaltered, the
Jastrow  function can be incorporated into $\ch(r_{ab}, \overline{E})$
giving an effective $\ch_{eff}(r_{ab}, \overline{E})$,
\be
\ch(r_{ab}, \overline{E}) \to 
\ch_{eff}(r_{ab}, \overline{E}) = \ch(r_{ab}, \overline{E}) 
[ 1 - \gamma_3 \;e^{-\gamma_1 \; r_{ab}^2} ( 1 - \gamma_2 \; r_{ab}^2 ) 
]^2 \;.
\label{eq.dbd3b}
\ee 
The standard set of values for the parameters $\gamma_1$, $\gamma_2$ and 
$\gamma_3$ are given ahead. The most important point about  Eq.
(\ref{eq.dbd1}), as mentioned earlier,   is that an experimental value of
(bound on) $T_{1/2}^{0\nu}$ will give a value for (bound on) neutrino mass
via Eq. (\ref{eq.dbd1}) provided we know the value of the square of NTME  
$M^{0\nu} (0^+)$ of the NDBD two-body transition operator $\co(2:0\nu)$,
connecting the ground state of the initial and final even-even nuclei 
involved.

Let us say that for the nuclei under consideration, protons are in the
single particle (sp) orbits $j^p$ and similarly neutrons in $j^n$. Using
the usual assumption that the radial part of the sp states are those of the
harmonic oscillator, the proton sp states are completely specified by
($\cn^p,\ell^p,j^p$) with $\cn^p$ denoting oscillator radial quantum number
so that for a oscillator shell $\can^p$, $2\cn^p+\ell^p=\can^p$. Similarly
the neutron sp states are ($\cn^n,\ell^n,j^n$). In terms of the creation
($a^\dagger$) and annihilation ($a$) operators, normalized two-particle
(antisymmetrized) creation operator $A^J_\mu(j_1j_2) = 
(1+\delta_{j_1j_2})^{-1/2}  (a^\dagger_{j_1}a^\dagger_{j_2})^J_\mu$ and
then $A^J_\mu \l|0\ran = \l|(j_1 j_2)J \mu\ran$ is a normalized 
two-particle state. With these the NDBD  transition operator  can be
written as,
\be
\co(2:0\nu) = \dis\sum_{j_1^p \geq j_2^p;j_3^n \geq j_4^n;J} \co_{j_1^p \, 
j_2^p; j_3^n \, j_4^n}^J (0\nu) \dis\sum_\mu A^J_\mu(j_1^pj_2^p) \l\{
A^J_\mu(j_3^nj_4^n) \r\}^\dagger \;.
\label{eq.dbd4}
\ee  
Note that $\co_{j_1^p \, j_2^p; j_3^n \, j_4^n}^J (0\nu) = \lan (j_1^p \, 
j_2^p) J M \mid \co(2:0\nu)  \mid (j_3^n \, j_4^n) J M \ran_a$ are two-body
matrix elements (TBME) and \lq{$a$}\rq denotes that we are considering 
antisymmetrized two-particle wavefunctions; $J$ is even  for $j_1=j_2$ 
or $j_3=j_4$. Numerical values for the TBME 's  in Eq. (\ref{eq.dbd4}) 
follow from the definition of $\co(2:0\nu)$ in Eq. (\ref{eq.dbd2}) 
which can be expressed as 
\be
\barr{c}
\co(2:0\nu)=\sum_{a,b} \tau_a^+ \tau_b^+ (\co^{GT} - (g_V^2/g_A^2)\co^F)
\;; \\
\\
\co^{GT} = \sigma_a \cdot \sigma_b \ch_{eff}(r_{ab},\overline{E})\;,\;\;\;
\co^F = \ch_{eff}(r_{ab},\overline{E}) \;.
\earr \label{eq.dbd4a} 
\ee
Note that $\tau_a^+ \tau_b^+$ simply changes two neutrons to two protons
and for the remaining parts, to obtain the TBME, we use Brody-Moshinsky 
brackets \cite{Mo-59,Br-60,Br-60a}; see \cite{Ba-09} for an alternative 
approach. Then, the TBME are given by,
\be
\barr{l}
\lan (j_1^p \, j_2^p) J M \mid \co(2:0\nu) \mid (j_3^n \, j_4^n) J M 
\ran_{a} =
\dis\frac{1}{\sqrt{(1+\delta_{j_1^pj_2^p})(1+\delta_{j_3^nj_4^n})}} \\ \\
\times \dis\sum_{L,S}  \l[2S(S+1)-3-\dis\frac{g_V^2}{g_A^2}\r] 
\chi \l\{ \begin{array}{ccc}
\ell_1^p & \ell_2^p & L \\
\spin & \spin & S \\
j_1^p & j_2^p & J \end{array} \r\} \chi \l\{ \begin{array}{ccc}
\ell_3^n & \ell_4^n & L \\
\spin & \spin & S \\
j_3^n & j_4^n & J \end{array} \r\} \\ \\
\times \dis\sum_{n,\ell;N,L^\pr} \l[ 1+(-1)^{\ell+S} \r]
\lan n \ell, N L^\pr, L \mid 
\cn_1^p \ell_1^p, \cn_2^p \ell_2^p, L \ran 
\times \lan n^\pr \ell, N L^\pr, L 
\mid \cn_3^n \ell_3^n, \cn_4^n \ell_4^n, L \ran \\ \\
\times \dis\sum_p B(n\ell, n^\pr\ell,p) \; I_p \;.
\earr\label{eq.dbd18}
\ee
Here, $\chi\{---\}$ are the 9$j$-coefficients, $\lan \cdots \mid \cdots
\ran$ are Brody-Moshinsky brackets \cite{Mo-59,Br-60,Br-60a},  $B(\cdots)$
are Brody, Jacob and Moshinsky coefficients \cite{Br-60} and $I_p$ are Talmi
integrals \cite{Br-60a}. It is important to mention that antisymmetrized
matrix elements for $\co(2:0\nu)$ are used in the shell model calculations
of NTME  while in QRPA related studies non-antisymmetrized matrix elements
are employed \cite{Ho-10,Si-08}.

There are a number of parameters in the NDBD transition operator and some of
them are usually varied in the calculations. The parameters are:  (i)
$R = 1.1 A^{1/3} - 1.2 A^{1/3}$ fm \cite{Si-09}; (ii) $b=1.003A^{1/6}$ fm
\cite{Ba-09}; (iii) $\overline{E}=1.12A^{1/2}$ MeV \cite{To-91}; (iv) 
$g_A/g_V=1$ (quenched) or 1.25 (unquenched) \cite{Si-09}; (v) three
different choices for the parameters  ($\gamma_1$, $\gamma_2$, $\gamma_3$)
in Eq. (\ref{eq.dbd3b}) are $(1.1,0.68,1)$ [Miller-Spencer],
$(1.52,1.88,0.46)$ [CD-Bonn] and $(1.59, 1.45, 0.92)$ [AV18]; these values
are taken from  \cite{Ho-10,Si-09}. It is useful to mention that the
kinematical factor $G^{0\nu}$ depends on the coupling constant $g_A$ 
(standard value is 1.25) and also some calculations use different values for
$r_0$ in $R=r_0 A^{1/3}$ fm; the standard value is $r_0=1.2$. Then a scaling
for $M^{0\nu}$ is \cite{Si-09,Me-11}
\be
\l(M^{0\nu}\r)^\prime = \l(\dis\frac{g_A}{1.25}\r)^2
\l(\dis\frac{1.2}{r_0}\r) M^{0\nu} \;.
\label{eq.m0nu}
\ee
Now we will give the spectral distribution formulation for calculating NTME
for NDBD.

\section{Spectral distribution method for NTME}
\label{sdm}

Following Eqs. (\ref{eq.dbd1}) and (\ref{eq.dbd2}) for NDBD half-lives, the
corresponding NTME $\l|M^{0\nu}\r|^2$ can be viewed as a transition 
strength (matrix element connecting a given initial state to a final state
by a transition operator) generated by the two-body transition  operator
$\co(2: 0\nu)$. Therefore, spectral distribution theory
\cite{Fr-88,Fr-88a,Fr-88b,Ko-01,Ko-01a}, based on random matrix theory,  for
transition strength densities, can be applied \cite{Ko-08}.  Transition
strength density is defined as the  transition strength multiplied by the
state densities at the initial and final energies involved. 

Let us consider shell model spherical orbits with angular momenta $j^p_1, 
j^p_2, \ldots, j^p_r$ with $m_p$ protons distributed over them and similarly
$m_n$ neutrons over $j^n_1, j^n_2, \ldots, j^n_s$ orbits. Then the proton
configurations $\tmp=[m_p^1, m_p^2, \ldots, m_p^r]$ where $m_p^i$ is number
of protons in the orbit $j_i^p$ and $\sum_{i=1}^r\,m_p^i=m_p$. Similarly the
neutron configurations $\tmn=[m_n^1, m_n^2, \ldots, m_n^s]$ where $m_n^i$ is
number of neutrons in the orbit $j_i^n$ and  $\sum_{i=1}^s\,m_n^i=m_n$. 
With these, $(\tmp, \tmn)$'s denote proton-neutron configurations.   The
nuclear effective Hamiltonian is one plus two-body, $H={\bf h} + {\bf V}$
and we  assume that the one-body part ${\bf h}$ includes the mean-field
producing part of the two-body part. Thus \cite{Fr-06,Fr-06a}, ${\bf V}$ is
the irreducible two-body part of $H$. The state density $I^H(E)$, with $\lan
\lan -- \ran \ran$ denoting trace, can be written as a sum of partial
densities defined over $(\tmp, \tmn)$, i.e.  $I^{(m_p, m_n)}(E) = \lan \lan
\delta(H-E) \ran \ran^{(m_p, m_n)} = \sum_{(\tmp, \tmn)}\, \lan \lan
\delta(H-E) \ran \ran^{(\tmp, \tmn)} = \sum_{(\tmp, \tmn)}\,I^{(\tmp,
\tmn)}(E) = \sum_{(\tmp, \tmn)}\,d(\tmp, \tmn)\,\rho^{(\tmp, \tmn)}(E)$.
Here $d(\tmp,\tmn)$ is dimension and  $\rho^{(\tmp, \tmn)}(E)$ is the
partial density normalized to unity. For nuclear Hamiltonians,  it is well 
understood \cite{Br-81, Ko-01, KH-10} that the smoothed form for the
eigenvalue densities is generated  by the action (locally) of EGOE(1+2) 
[embedded GOE of one plus two-body interactions] operating in  the Gaussian
domain. This gives, 
\be
I^{(m_p, m_n)}(E) = \dis\sum_{(\tmp, \tmn)}\;I^{(\tmp, \tmn)}_\cg(E)\;.
\label{eq.13}
\ee
In Eq. (\ref{eq.13}), $\cg$ denotes Gaussian. The Gaussian partial
densities are defined by the centroids $E_c(\tmp, \tmn)=\lan H
\ran^{(\tmp, \tmn)} \sim \lan {\bf h} \ran^{(\tmp, \tmn)}$ and variances
$\sigma^2(\tmp, \tmn) = \lan H^2  \ran^{(\tmp, \tmn)} -
[E_c(\tmp,\tmn)]^2$ $\sim \lan {\bf V}^2 \ran^{(\tmp, \tmn)}$ and, as they
are traces over  $(\tmp, \tmn)$ spaces, they can be calculated without
recourse to $H$ matrix construction. Propagation equations for them, in
terms of the sp energies defining ${\bf h}$ and TBME defining ${\bf V}$,
are easy to write down.  Unitary group tensorial structure of ${\bf h}$
and ${\bf V}$ gives further simplifications of Eq. (\ref{eq.13}); see
\cite{Fr-06,Fr-06a,Km-96} for details and applications to $fp$-shell 
and also for heavy $(A \gazz 150)$ nuclear data analysis.

Random matrix theory, based on EGOE(1+2), for the (smoothed)  transition
strength densities $I_\co(E_i, E_f) = I(E_f) |\lan E_f \mid \co \mid E_i
\ran|^2 I(E_i)$ allows us to write $I_\co$ as a convolution of the
corresponding   density generated by the mean-field part ${\bf h}$ with a
spreading bivariate Gaussian $\rho_{biv-\cg :  \co : {\bf V}}$ due to the
interaction ${\bf V}$.  This gives \cite{Ko-01,Fr-88,Fr-88a,Fr-88b}, for the
square of the $\co$ matrix elements, with $\l|0^+_i\ran = \l|E_i\ran$ and 
$\l|0^+_f\ran = \l|E_f\ran$ where $E$'s are energies,
\be
\barr{l}
\l|\lan E_f \mid \co \mid E_i \ran\r|^2 = 
\dis\sum_{(\tmp, \tmn)_i,\,(\tmp, \tmn)_f}\; 
\dis\frac{I^{(\tmp, \tmn)_i}_\cg
(E_i) \; I^{(\tmp, \tmn)_f}_\cg (E_f)} {I^{(m_p, m_n)_i} (E_i) \; 
I^{(m_p, m_n)_f}(E_f)}
\\ \\
\times \l|\lan (\tmp, \tmn)_f \mid \co \mid 
(\tmp, \tmn)_i \ran\r|^2  
\; \dis\frac{\rho^{(\tmp, \tmn)_i,\,(\tmp, \tmn)_f}_{biv-\cg : 
\co : {\bf V}} (E_i,E_f\;;\; 
E_c^i, E_c^f, \sigma_i, \sigma_f, \zeta)}{\rho_{\cg}^{(\tmp,
\tmn)_i}(E_i)\;\rho_{\cg}^{(\tmp, \tmn)_f}(E_f)} \;.
\earr \label{eq.14}
\ee
In Eq. (\ref{eq.14}), $\l|\lan (\tmp, \tmn)_f \mid \co
\mid (\tmp, \tmn)_i \ran \r|^2$ is the mean square matrix element of
$\co$ connecting  $(\tmp, \tmn)_i$ and $(\tmp, \tmn)_f$
configurations,
\be
\barr{rcl}
\l|\lan (\tmp, \tmn)_f \mid \co \mid (\tmp, \tmn)_i 
\ran\r|^2 & = & \l\{d[(\tmp, \tmn)_i]\,d[(\tmp, \tmn)_f]\r\}^{-1}  
\\ \\
& \times &
\dis\sum_{\alpha \in 
(\tmp, \tmn)_i, \;\beta \in (\tmp, \tmn)_f}\;\l|\lan (\tmp, \tmn)_f \;
\beta \mid \co \mid (\tmp, \tmn)_i \;\alpha \ran\r|^2\;.
\earr \label{eq.15}
\ee
For later reference it is also useful define,
\be
\lan [\co]^\dagger \co \ran^{(\tmp, \tmn)_i} 
= d[(\tmp, \tmn)_f] \;\; \dis\sum_{(\tmp, \tmn)_f} 
\l|\lan (\tmp, \tmn)_f \mid \co \mid (\tmp,\tmn)_i \ran\r|^2 \;.
\label{eq.16}
\ee
Following \cite{Fr-88,Fr-88a,Fr-88b}, the two centroids $E_c^i$
and $E_c^f$ and the two variances $\sigma^2_i$ and $\sigma_f^2$ of the
marginal densities of the bivariate Gaussian $\rho_{biv-\cg : \co: 
{\bf V}}$, with some approximations are given by,
\be
\barr{c}
E_c^i=E_c\l((\tmp, \tmn)_i\r)\;,\;\;\;\;E_c^f=E_c\l((\tmp, \tmn)_f\r)
\;, \\ \\
\sigma^2_i = \sigma^2\l((\tmp, \tmn)_i\r)\;,\;\;\;\;
\sigma^2_f = \sigma^2\l((\tmp, \tmn)_f\r)\;.
\earr \label{eq.17a}
\ee
These are the proton-neutron configuration centroids and variances generated
by $H$. Eq. \eqref{eq.17a} has its basis in Eq. \eqref{eq.9ab} given ahead.
Although the general structure and importance of the  bivariate correlation
coefficient $\zeta$ in Eq. (\ref{eq.14}) is well  understood
\cite{Fr-88,Fr-88a,Fr-88b}, an expression for $\zeta$ in terms of traces 
over $(\tmp, \tmn)_i$ and $(\tmp, \tmn)_f$ spaces is not yet available. Then
the standard approximation, for completing the theory, is
\be
\barr{c}
\zeta = \dis\frac{X_{11}\l[(m_p,m_n)_i\r]}{\dis\sqrt{X_{20}\l[(m_p,m_n)_i
\r]\;X_{02}\l[(m_p,m_n)_i\r]}}\;; \;\;
X_{PQ}\l[(m_p,m_n)_i\r] =
\dis\frac{\lan [\co]^\dagger {\bf V}^Q \co {\bf V}^P \ran^{(m_p,m_n)_i}}
{\lan [\co]^\dagger \co \ran^{(m_p,m_n)_i}} \;. 
\earr \label{eq.18}  
\ee
To proceed further, propagation formulas for the traces in  Eqs.
\eqref{eq.15} and \eqref{eq.18} are needed.  For the trace
in Eq. (\ref{eq.15}), using the results in \cite{CFT},  it  is easy to write
the propagation formula in terms of the  TBME $\co^{J_0}_{--}(0\nu)$ defined
by Eq. (\ref{eq.dbd4}),
\be
\barr{l}
\l|\lan (\tmp, \tmn)_f \mid \co \mid (\tmp, 
\tmn)_i \ran\r|^2 \; d[(\tmp, \tmn)_f] 
\\ \\
= \;\dis\sum_{\alpha , \beta , \gamma ,\delta}\;
\dis\frac{m^i_n(\alpha) [m^i_n(\beta)-\delta_{\alpha \beta}] 
[N_p(\gamma) - m^i_p(\gamma)] [N_p(\delta) -
m^i_p(\delta) - \delta_{\gamma \delta}]}{N_n(\alpha) [N_n(\beta) -
\delta_{\alpha \beta}] N_p(\gamma) [N_p(\delta) - \delta_{\gamma \delta}]}
\\ \\
\times \dis\sum_{J_0} \; \l[\co^{J_0}_{\gamma^p \delta^p \alpha^n
\beta^n}(0\nu)\r]^2
(2J_0 +1) \;,\\ \\
(\tmp, \tmn)_f = (\tmp, \tmn)_i \times \l(1^+_{\gamma_p} 1^+_{\delta_p}
1_{\alpha_n} 1_{\beta_n} \r) \;.
\earr \label{eq.19}
\ee
Note that in Eq. (\ref{eq.19}) the final configuration is defined by
removing  one neutron from orbit $\alpha$ and another from $\beta$ and then
adding one proton in orbit $\gamma$ and another in orbit $\delta$. Also,
$N_p(\alpha)$ is the degeneracy of the proton orbit $\alpha$ and similarly
$N_n(\alpha)$ for the neutron orbit $\alpha$. In the limit $\rho_{biv}/\rho
\rho = 1$ in Eq. (\ref{eq.14}), substituting the result of Eq. (\ref{eq.19})
in Eq. (\ref{eq.14}) gives the NTME for NDBD. However, this zero-th order
approximation will not be good as in general it is expected that $\zeta > 
0.5$; see Sec. V ahead. Before proceeding to implement the theory given
above, it is essential to test the important approximation used in the
theory, i.e. the bivariate Gaussian form for the spreading function
generated by ${\bf V}$ by calculating the fourth order bivariate cumulants
(they will be zero for a bivariate Gaussian). In addition we also need
an expression for the bivariate correlation coefficient $\zeta$.  To 
address these two problems and provide generic results, we consider the 
spreading function defined over proton-neutron spaces i.e., 
$$
\rho^{(m_p, m_n)_i,\,(m_p, m_n)_f}_{biv : 
\co : H} (E_i,E_f\;;\; E_c^i, E_c^f, \sigma_i, \sigma_f, \zeta) \;,
$$
where $H$ is a two-body Hamiltonian. We will consider Hamiltonians that
preserve $(m_p,m_n)$ and then $H = H_{pp} + H_{nn} + H_{pn}$. This is quite
appropriate for heavy nuclei. Adopting the binary correlation theory, the
bivariate moments of $\rho^{(m_p, m_n)_i,\,(m_p, m_n)_f}_{biv : \co : H}$
are evaluated by considering random $k$-body $H$ operators. Similarly, we
represent the transition operator  $\co$ by random $k_\co$-body operator
that changes $k_\co$ number of neutrons to $k_\co$ number of neutrons. This
is equivalent to using EGOE representation for both $H$ and $\co$ operators 
\cite{FKPT,Ko-01}. Let us mention that, from now onwards, we consider only
the two-orbit configurations $(m_p,m_n)$ [for generality, these are denoted
as $(m_1,m_2)$ in the next two sections].

\section{Binary correlation results for random Hamiltonians}
\label{c7s1}

Binary correlation theory for moments defined over a single unitary orbit is
given by Mon and French \cite{Mo-73,MF-75} and they correspond to the
moments generated by spinless EGOE$(k)$ in the dilute limit (dilute limit is
defined in Sec. IV). The theory is extended to certain two-orbit moments by
Tomsovic \cite{To-86}. For the two problems mentioned in Sec. III,  we need
traces defined over two-orbits (protons and neutrons) with the $H$
preserving the two-orbit symmetry and the transition operator $\co$ acting
on a two-orbit configuration generating a unique final two-orbit
configuration. In the present section, we will give the basic binary
correlation results adopted for this situation and in Sec. V, we will
consider their applications. 

\subsection{Results for single unitary orbit}

Let us begin with a $k_H$-body operator,
\be
H(k_H) = \dis\sum_{\alpha,\; \beta} v_H^{\alpha\beta}\; \alpha^\dg(k_H)
\beta(k_H) \;.
\label{eq.b1}
\ee
Here, $\alpha^\dg(k_H)$ is the $k_H$ particle creation operator and
$\beta(k_H)$ is the $k_H$ particle annihilation operator. Similarly,
$v_H^{\alpha\beta}$ are matrix elements of the operator $H$ in $k_H$
particle space i.e., $v_H^{\alpha\beta} = \lan k_H \beta \mid H \mid k_H 
\alpha \ran$ (it should be noted that Mon and French \cite{Mo-73,MF-75} 
used operators with daggers to denote annihilation operators and operators
without daggers to denote creation operators). Following basic traces will
be used throughout,
\be
\dis\sum_\alpha \alpha^\dg(k) \alpha(k)  = \dis\binom{\hat{n}}{k} \;\;\;\;
\Rightarrow \lan \dis\sum_\alpha \alpha^\dg(k) \alpha(k) \ran^m =  
\dis\binom{m}{k} \;.
\label{eq.b2}
\ee
\be
\dis\sum_\alpha \alpha(k) \alpha^\dg(k) = \dis\binom{N - \hat{n}}{k} 
\;\;\;\;
\Rightarrow \lan \dis\sum_\alpha \alpha(k) \alpha^\dg(k) \ran^m =  
\dis\binom{\wm}{k} \;; \;\;\;\; \wm = N-m \;.
\label{eq.b3}
\ee
\be
\barr{l}
\dis\sum_\alpha \alpha^\dg(k) B(k^\pr) \alpha(k) = 
\dis\binom{\hat{n} - k^\pr}{k} B(k^\pr) \\ \\
\Rightarrow \lan \dis\sum_\alpha \alpha^\dg(k) B(k^\pr) \alpha(k) \ran^m =  
\dis\binom{m-k^\pr}{k} B(k^\pr) \;.
\earr \label{eq.b4}
\ee
\be
\barr{l}
\dis\sum_\alpha \alpha(k) B(k^\pr) \alpha^\dg(k) = 
\dis\binom{N - \hat{n} - k^\pr}{k} B(k^\pr) \\ \\
\Rightarrow \lan \dis\sum_\alpha \alpha(k) B(k^\pr) \alpha^\dg(k) \ran^m =  
\dis\binom{\wm-k^\pr}{k} B(k^\pr) \;.
\earr \label{eq.b5}
\ee
Equation (\ref{eq.b2}) follows from the fact that the average should be zero
for $m < k$ and one for $m=k$ and similarly, Eq. (\ref{eq.b3})  follows from
the same argument except that the particles are replaced by holes. Equation
(\ref{eq.b4}) follows first by writing the $k^\pr$-body operator $B(k^\pr)$
in operator form using Eq. (\ref{eq.b1}), i.e.,
\be
B(k^\pr) = \dis\sum_{\beta,\; \gamma} v_B^{\beta\gamma}\; \beta^\dg(k^\pr)
\gamma(k^\pr) \;,
\label{eq.b5a}
\ee
and then applying the commutation relations for the fermion creation and
annihilation operators. This gives $\sum_{\beta,\gamma} v_B^{\beta\gamma}\;
\beta^\dg(k^\pr)  \sum_\alpha \alpha^\dg(k) \alpha(k) \gamma(k^\pr)$. Now
applying Eq. (\ref{eq.b2}) to the sum involving $\alpha$ gives Eq.
(\ref{eq.b4}). Eq. (\ref{eq.b5}) follows from the same arguments except one
has to assume that $B(k^\pr)$ is fully irreducible $\nu = k^\pr$ operator
and therefore, it has particle-hole symmetry. For  a general $B(k^\pr)$
operator, this is valid only in the $N \to \infty$ limit. Therefore, this
equation has to be  applied with caution.

Using the definition of the $H$ operator in Eq. (\ref{eq.b1}), 
we have
\be
\barr{rcl}
\overline{\lan H(k_H) H(k_H) \ran^m} & = & \dis\sum_{\alpha,\;\beta}
\overline{\l\{v_H^{\alpha\beta}\r\}^2} \; \lan \alpha^\dg(k_H) \beta(k_H)
\beta^\dg(k_H) \alpha(k_H) \ran^m \\ \\
& = & v_H^2 \; \lan \dis\sum_\alpha \alpha^\dg(k_H) \l\{ 
\dis\sum_\beta \beta(k_H) \beta^\dg(k_H) \r\} \alpha(k_H) \ran^m
\\ \\
& = & v_H^2 \; T(m,N,k_H) \;.
\earr \label{eq.b6a1}
\ee
Here, $H$ is taken as EGOE$(k_H)$ with all the $k_H$ particle matrix
elements being Gaussian variables with zero center and same variance for
off-diagonal matrix elements (twice for the diagonal matrix elements). This
gives  $\overline{(v_H^{\alpha \beta})^2} = v_H^2$ to be independent of 
$\alpha, \; \beta$ labels. It is important to note that in the dilute limit,
the diagonal terms [$\alpha =\beta$ in Eq. (\ref{eq.b6a1})] in the averages
are neglected (as they are smaller by at least one power of $1/N$)  and the
individual $H$'s are unitarily irreducible. These assumptions are no longer
valid for finite-$N$ systems and hence, evaluation of averages is more
complicated. In the dilute limit, we have
\be
\barr{rcl}
T(m,N,k_H) & = &  \lan \dis\sum_\alpha \alpha^\dg(k_H) \l\{ 
\dis\sum_\beta \beta(k_H) \beta^\dg(k_H) \r\} \alpha(k_H) \ran^m
\\ \\
& = &  \dis\binom{\wm+k_H}{k_H} \; \lan \dis\sum_\alpha \alpha^\dg(k_H)
\alpha(k_H) \ran^m \\ \\ 
& = &  \dis\binom{\wm+k_H}{k_H} \; \dis\binom{m}{k_H} \;.
\earr \label{eq.b6a}
\ee
Note that, we have used Eq. (\ref{eq.b3}) to evaluate the summation over
$\beta$ and Eq. (\ref{eq.b2}) to evaluate summation over $\alpha$ in Eq.
(\ref{eq.b6a}). In the `strict' $N \to \infty$ limit, we have
\be
\barr{l} 
T(m,N,k_H) \stackrel{N \to \infty}{\to} \dis\binom{m}{k_H} \; 
\dis\binom{N}{k_H} \;.
\earr \label{eq.b6b}
\ee
In order to incorporate the finite-$N$ corrections, we have to consider the
contribution of the diagonal terms. Then, we have,
\be
T(m,N,k_H) = \dis\binom{m}{k_H}\l[ \dis\binom{\wm+k_H}{k_H} + 1 \r]\;.
\label{eq.5b}
\ee
Now we will turn to the fourth order averages.

For averages involving product of four operators of the form $\lan H(k_H)
G(k_G) H(k_H) G(k_G) \ran^m$,  with operators $H$ and $G$ independent and of
body ranks $k_H$ and $k_G$ respectively, there are two possible ways of
evaluating this trace. Either (a) first contract the $H$ operators across
the $G$ operator using Eq. (\ref{eq.b5}) and  then contract the $G$
operators  using Eq. (\ref{eq.b4}),  or (b) first contract the $G$ operators
across the $H$ operator  using Eq. (\ref{eq.b5}) and then contract the $H$
operators using Eq. (\ref{eq.b4}). However, (a) and (b) give the same
result only in the `strict'  $N \to \infty$ limit and also for the result
incorporating finite $N$ corrections as discussed below. In general, the
final result can be expressed as,
\be
\overline{\lan H(k_H) G(k_G) H(k_H) G(k_G) \ran^m} = 
v_H^2 \; v_G^2 \; F(m,N,k_H,k_G)\;.
\label{eq.b7}
\ee
In the `strict' dilute  limit, $F(m,N,k_H,k_G)$ is given by
\be
F(m,N,k_H,k_G) =  \dis\binom{m-k_H}{k_G} \; \dis\binom{m}{k_H} \; 
\dis\binom{N}{k_H} \; \dis\binom{N}{k_G}\;.
\label{eq.b8}
\ee
In order to obtain finite-$N$ corrections to $F(\cdots)$, we have to
contract over operators whose lower symmetry parts can not be ignored. The
operator $H(k_H)$ decomposes into irreducible symmetry parts $\cf(s)$
denoted by  $s=0,1,2,\ldots,k_H$ with respect to the unitary group $SU(N)$.
For a $k_H$-body number conserving operator \cite{CFT,MF-75}, we have
\be
H(k_H) = \dis\sum_{s=0}^{k_H} \; \dis\binom{m-s}{k_H-s} \; \cf(s) \;.
\label{eq.b8a}
\ee 
Here, the $\cf(s)$ are orthogonal with respect to $m$-particle averages, 
i.e., $\lan \cf(s) \cf^\dg(s^\pr) \ran^m = \delta_{ss^\pr}$. Now,
${\lan  H(k_H) G(k_G) H(k_H) G(k_G) \ran^m}$ will have four parts,
\be
\barr{l}
\overline{\lan  H(k_H) G(k_G) H(k_H) G(k_G) \ran^m} \\ \\
= v_H^2 v_G^2
\dis\sum_{\alpha,\beta,\gamma,\delta} \lan \alpha^\dg(k_H) \beta(k_H)
\gamma^\dg(k_G)
\delta(k_G)  \beta^\dg(k_H) \alpha(k_H) \delta^\dg(k_G) \gamma(k_G) \ran^m 
\\ \\
+ v_H^2 v_G^2
\dis\sum_{\alpha,\gamma,\delta} \lan \alpha^\dg(k_H) 
\alpha(k_H) \gamma^\dg(k_G)
\delta(k_G)  \alpha^\dg(k_H) \alpha(k_H) \delta^\dg(k_G) \gamma(k_G) \ran^m 
\earr \label{eq.2}
\ee
\be
\barr{l}
+ v_H^2 v_G^2
\dis\sum_{\alpha,\beta,\gamma} \lan \alpha^\dg(k_H) 
\beta(k_H) \gamma^\dg(k_G)
\gamma(k_H)  \beta^\dg(k_H) \alpha(k_H) \gamma^\dg(k_G) \gamma(k_G) \ran^m
\\ \\
+ v_H^2 v_G^2
\dis\sum_{\alpha,\delta} \lan \alpha^\dg(k_H) \alpha(k_H) \delta^\dg(k_G)
\delta(k_G)  \alpha^\dg(k_H) \alpha(k_H) \delta^\dg(k_G) \delta(k_G) \ran^m 
\\ \\
= X + Y_1 + Y_2 + Z \;. \nonumber
\earr \label{eq.2-1}
\ee
Note that we have decomposed each operator into diagonal and off-diagonal
parts. We have used the condition that the variance of the diagonal matrix
elements is  twice that of the off-diagonal matrix elements in the defining
spaces to convert the restricted  summations into unrestricted summations
appropriately to obtain the four terms in the  RHS of Eq. (\ref{eq.2}).
Following  \cite{Mo-73,To-86,Ma-11}  and applying unitary decomposition to
$\gamma \delta^\dagger$ (also $\delta \gamma^\dagger$) in the first two
terms and $\alpha \beta^\dagger$ (also $\beta \alpha^\dagger$) in the  third
term we get $X$, $Y_1$ and $Y_2$. To make things clear, we will discuss the
derivation for $X$ term in detail before proceeding further. Applying
unitary decomposition to the operators $\gamma^\dg(k_G)\delta(k_G)$ and 
$\gamma(k_G)\delta^\dg(k_G)$ using Eq. (\ref{eq.b8a}), we have
\be
X = v_H^2 \; v_G^2 \; \dis\sum_{\alpha,\beta,\gamma,\delta} \;
\dis\sum_{s=0}^{k_G} \; \dis\binom{m-s}{k_G-s}^2 \; 
\lan \alpha^\dg(k_H) \beta(k_H) \cf^\dg_{\gamma\delta}(s) 
\beta^\dg(k_H) \alpha(k_H) \cf_{\gamma\delta}(s) \ran^m \;.
\label{eq.b8ba}
\ee
Contracting the operators $\beta\beta^\dg$ across $\cf$'s using Eq.
(\ref{eq.b5}) and operators $\alpha^\dg\alpha$ across $\cf$ using Eq.
(\ref{eq.b4}) gives,
\be
X
= v_H^2 \; v_G^2 \; 
\dis\sum_{s=0}^{k_G} \; \dis\binom{m-s}{k_G-s}^2 \; 
\dis\binom{\wm+k_H-s}{k_H}
\; \dis\binom{m-s}{k_H} \; \dis\sum_{\gamma,\delta} \;
\lan \cf^\dg_{\gamma\delta}(s) \cf_{\gamma\delta}(s) \ran^m \;.
\label{eq.b8b}
\ee
Inversion of the equation,
\be
\dis\sum_{\gamma,\delta} 
\lan \gamma^\dg(k_G) \delta(k_G) \delta^\dg(k_G) \gamma(k_G) \ran^m 
= Q(m) = \dis\sum_{s=0}^{k_G} 
\dis\binom{m-s}{k_G-s}^2 \; \dis\sum_{\gamma,\delta} 
\lan \cf^\dg_{\gamma\delta}(s) \cf_{\gamma\delta}(s) \ran^m \;,
\label{eq.b8c}
\ee
gives,
\be
\barr{l}
\dis\binom{m-s}{k_G-s}^2 \; \dis\sum_{\gamma,\delta} 
\lan \cf^\dg_{\gamma\delta}(s) \cf_{\gamma\delta}(s) \ran^m 
= 
\dis\binom{m-s}{k_G-s}^2 \; \dis\binom{N-m}{s} \; \dis\binom{m}{s}\;
\l[ (k_G-s)! s! \r]^2   
\\ \\
\times (N-2s+1) \;
\dis\sum_{t=0}^s \dis\frac{(-1)^{t-s}\l[ (N-t-k_G)!\r]^2}
{(s-t)! (N-s-t+1)! t! (N-t)!} Q(N-t) \;.
\earr \label{eq.b8d}
\ee
For the average required in Eq. (\ref{eq.b8c}), we have
\be
Q(m) = \dis\sum_{\gamma,\delta} \lan \gamma^\dg(k_G) 
\delta(k_G) \delta^\dg(k_G)
\gamma(k_G) \ran^m = \dis\binom{\wm+k_G}{k_G} \; \dis\binom{m}{k_G} \;.
\label{eq.b8e}
\ee
Simplifying Eq. (\ref{eq.b8d}) using Eq. (\ref{eq.b8e}) and using the 
result in Eq. (\ref{eq.b8b}) along with the series summation \cite{Mo-73}
\be
\dis\sum_{t=0}^s \;\dis\frac{(-1)^{t-s} (N-t-k_G)! \; (k_G+t)!}
{(s-t)! \; (t!)^2 \; (N-s-t+1)!} = \dis\frac{k_G! (N-k_G-s)!}{(N+1-s)!} \;
\dis\binom{k_G}{s} \;\dis\binom{N+1}{s} \;,
\label{eq.ser-x}
\ee
the expression for $X$ is,
\be
\barr{rcl}
X & = & v_H^2 v_G^2 \; F(m,N,k_H,k_G) \;;  \\ \\
F(m,N,k_H,k_G) & = & \dis\sum_{s=0}^{k_G} \dis\binom{m-s}{k_G-s}^2
\dis\binom{\wm+k_H-s}{k_H} \dis\binom{m-s}{k_H} \dis\binom{\wm}{s}
\dis\binom{m}{s} \dis\binom{N+1}{s} \\ \\
& \times & \dis\frac{N-2s+1}{N-s+1}
\dis\binom{N-s}{k_G}^{-1} \dis\binom{k_G}{s}^{-1} \;.
\earr \label{eq.3}
\ee
Although not obvious, $X$ has $k_H \leftrightarrow k_G$ symmetry. This is
easy to verify for $k_H, k_G \leq 2$. In the large $N$ limit,
$Y_1$, $Y_2$ and $Z$ are neglected as $X$ will make the dominant
contribution; see \cite{Ma-11} for details on $Y_1$, $Y_2$ and $Z$. Thus, in
all the applications, we use 
\be
\overline{\lan  H(k_H) G(k_G) H(k_H) G(k_G) \ran^m} = X = 
v_H^2 \; v_G^2 \;  F(m,N,k_H,k_G) \;.
\label{eq.6}
\ee
An immediate application of these averages is in evaluating the fourth order
average $\overline{\lan H^4(k_H) \ran^m}$. 
There will be three different correlation
patterns that will contribute to this average in the binary correlation
approximation (we must correlate in pairs the operators for all moments 
of order $> 2$),
\be
\barr{rcl}
\overline{\lan H^4(k_H) \ran^m} & = & 
\overline{\bcon{}{H(k_H)}{}{H(k_H)}
\bcon{H(k_H) H(k_H)}{H(k_H)}{}{H(k_H)}
\lan H(k_H) H(k_H) H(k_H) H(k_H) \ran^m} \\ \\
& + &
\overline{\bcon{}{H(k_H)}{H(k_H)H(k_H)}{H(k_H)}
\bcon[2ex]{H(k_H)}{H(k_H)}{}{H(k_H)}
\lan H(k_H) H(k_H) H(k_H) H(k_H) \ran^m}
\\ \\ 
& + &
\overline{\bcon{}{H(k_H)}{H(k_H)}{H(k_H)}
\bcon[2ex]{H(k_H)}{H(k_H)}{H(k_H)}{H(k_H)}
\lan H(k_H) H(k_H) H(k_H) H(k_H) \ran^m}
\;.
\earr \label{eq.b10a}
\ee  
In Eq. \eqref{eq.b10a}, we denote the binary correlated pairs of operators
with the symbol $\bcon{}{H}{}{H} HH$. The first two terms on the RHS of Eq. (\ref{eq.b10a})
are equal due to cyclic invariance and follow from Eq. 
(\ref{eq.b6a1}),
\be
\barr{rcl}
\overline{\bcon{}{H(k_H)}{}{H(k_H)}
\bcon{H(k_H) H(k_H)}{H(k_H)}{}{H(k_H)}
\lan H(k_H) H(k_H) H(k_H) H(k_H) \ran^m} & = &
\overline{\bcon{}{H(k_H)}{H(k_H)H(k_H)}{H(k_H)}
\bcon[2ex]{H(k_H)}{H(k_H)}{}{H(k_H)}
\lan H(k_H) H(k_H) H(k_H) H(k_H) \ran^m} \\ \\
& = &
\l[ \; \overline{\lan H^2(k_H) \ran^m} \; \r]^2\;.
\earr \label{eq.b10b}
\ee
Similarly, the third term on the RHS of Eq. (\ref{eq.b10a}) follows  
from Eq. (\ref{eq.6}),
\be
\overline{\bcon{}{H(k_H)}{H(k_H)}{H(k_H)}
\bcon[2ex]{H(k_H)}{H(k_H)}{H(k_H)}{H(k_H)}
\lan H(k_H) H(k_H) H(k_H) H(k_H) \ran^m} = 
v_H^4 \; F(m,N,k_H,k_H) \;. 
\label{eq.b10c}
\ee
Combining Eqs. \eqref{eq.b10a}, \eqref{eq.b10b} and \eqref{eq.b10c}, 
$\overline{\lan H^4(k_H) \ran^m}$ is given by,
\be
\overline{\lan H^4(k_H) \ran^m} =  v_H^4 \; \l[ 2 \; \l\{T(m,N,k_H)\r\}^2 + 
F(m,N,k_H,k_H) \r]\;.
\label{eq.b10}
\ee

\subsection{Results for two unitary orbits}

In the NDBD applications (also $\beta$ decay),  fourth order traces over two
orbit configurations are needed.  Let us consider $m$ particles in two
orbits with number of sp states being $N_1$ and $N_2$ respectively. Now the
$m$-particle space can be divided into configurations $(m_1,m_2)$ with $m_1$
particles in the \#1 orbit and $m_2$ particles in the \#2 orbit such that $m
= m_1 + m_2$. Considering the operator $H$ with fixed body ranks in $m_1$
and $m_2$ spaces such that $(m_1,m_2)$ are preserved by this operators, the
general form for $H$ is,
\be
H(k_H) = \dis\sum_{i+j=k_H\;;\alpha,\beta,\gamma,\delta} 
\l[ v_H^{\alpha\beta\gamma\delta}(i,j) \r] \; \alpha_1^\dg(i) 
\beta_1(i) \gamma_2^\dg(j)\delta_2(j) \;.
\label{eq.b11}
\ee
Now, it is seen that, in the dilute limit, 
\be
\barr{l}
\overline{\lan H^2(k_H) \ran^{m_1,m_2}} \\ \\
= 
\dis\sum_{i+j=k_H} v_H^2(i,j)\;
\dis\sum_{\alpha,\beta,\gamma,\delta} \lan \alpha_1^\dg(i) 
\beta_1(i) \gamma_2^\dg(j)\delta_2(j) \beta_1^\dg(i) 
\alpha_1(i) \delta_2^\dg(j)\gamma_2(j) \ran^{m_1,m_2}
\\ \\
=
\dis\sum_{i+j=k_H} v_H^2(i,j)\;
\dis\sum_{\alpha,\beta} \lan \alpha_1^\dg(i) 
\beta_1(i)  \beta_1^\dg(i) \alpha_1(i) \ran^{m_1} \dis\sum_{\gamma,\delta}
\lan \gamma_2^\dg(j)\delta_2(j) \delta_2^\dg(j)\gamma_2(j) \ran^{m_2} 
\\  \\
= \dis\sum_{i+j=k_H} v_H^2(i,j)\;
T(m_1,N_1,i)\;T(m_2,N_2,j) \;.
\earr \label{eq.b12}
\ee
Note that $v_H^2(i,j) = \overline{[v_H^{\alpha\beta\gamma\delta} (i,j)]^2}$
and $T$'s are defined by Eqs. (\ref{eq.b6a}) and (\ref{eq.b6b}). The
ensemble is defined such that $v_H^{\alpha\beta\gamma\delta}(i,j)$ are
independent Gaussian random variables with zero center and the variances
depend only on the indices $i$ and $j$.  Similarly, with two operators $H$
and $G$ (with body ranks $k_H$ and $k_G$ respectively) that are independent
and both preserving $(m_1,m_2)$, $\overline{\lan H(k_H) G(k_G) H(k_H) G(k_G)
\ran^{m_1,m_2}}$ is  given by,
\be
\barr{l}
\overline{\lan H(k_H) G(k_G) H(k_H) G(k_G) \ran^{m_1,m_2}} =  \\ \\
\dis\sum_{i+j=k_H,\; t+u=k_G} \; v_H^2(i,j) \; v_G^2(t,u) \;
F(m_1,N_1,i,t) \;  F(m_2,N_2,j,u) \;,
\earr \label{eq.b13}
\ee
and therefore,
\be
\barr{l}
\overline{\lan H^4(k_H) \ran^{m_1,m_2}}
= 2\; 
\l[ \dis\sum_{i+j=k_H} v_H^2(i,j)\;
T(m_1,N_1,i)\;T(m_2,N_2,j) \r]^2 \\ \\ 
+ 
\dis\sum_{i+j=k_H,\; t+u=k_H} \; v_H^2(i,j) \; 
v_H^2(t,u) \; F(m_1,N_1,i,t) \;  F(m_2,N_2,j,u) \;.
\earr \label{eq.b14}
\ee
Now we apply the formulation given here to derive the formulas for the
second and fourth order cumulants defining
$\rho_{biv-\cg: \co :H}^{(m_p,m_n)_i,  (m_p,m_n)_f}$.

\section{Binary correlation results for the bivariate correlation
coefficient and fourth order cumulants for the transition strength density
for NDBD}

\subsection{Transition matrix elements and bivariate strength densities}
\label{c7s3s1}

Our purpose here is to establish that for the $0\nu\beta^-\beta^-$ decay 
(also for $\beta$ decay), transition strength densities, locally, are 
close to bivariate Gaussian form  and also to derive a formula for the 
corresponding bivariate correlation coefficient. With space \#1 denoting  
protons and similarly space \#2 neutrons, the general form of the 
transition operator $\co$ is,
\be
\co(k_\co) = \dis\sum_{\gamma,\delta} v_\co^{\gamma\delta}(k_\co)\; 
\gamma_1^\dg(k_\co) \delta_2(k_\co) \;; \;\;\;\; k_\co=2 \; 
\mbox{for NDBD}\;.
\label{eq.9}
\ee
Therefore, in order to derive the form for the transition strength densities
generated by $\co$, it is necessary to deal with two-orbit configurations
denoted by $(m_1,m_2)$, where $m_1$ is the number of particles in the first
orbit (protons) and $m_2$ in the second orbit (neutrons).   Now, the
transition strength density  $I_\co(E_i,E_f)$ is
\be
\barr{l}
I_\co^{(m_1, m_2)_i, (m_1, m_2)_f}(E_i,E_f) \\ \\ =  
I^{(m_1, m_2)_f}(E_f) \l|\lan (m_1, m_2)_f E_f 
\mid \co \mid (m_1, m_2)_i E_i \ran \r|^2
I^{(m_1, m_2)_i}(E_i) \;,
\earr \label{eq.bbd3}
\ee
and the corresponding bivariate moments are  
\be
\wtM_{PQ}((m_1, m_2)_i) =   \overline{\lan \co^\dagger(k_\co)
H^Q(k_H) \co(k_\co) H^P(k_H) \ran^{(m_1, m_2)_i}} \;.
\label{eq.bbd4}
\ee
Note that $\wtM$ are in general non-central and non-normalized moments.  The
general form of the operator $H(k_H)$ is given by Eq. (\ref{eq.b11}) and it
preserves $(m_1, m_2)_i$'s.  However,  $\co$ and its hermitian conjugate
$\co^\dg$ do not preserve $(m_1,m_2)$ i.e.,   $\co(k_\co) \l| m_1, m_2 \ran
= \l| m_1 + k_\co, m_2 - k_\co \ran$ and  $\co^\dg(k_\co) \l| m_1, m_2 \ran
= \l| m_1 - k_\co, m_2 + k_\co \ran$. Thus, given a $(m_1, m_2)_i$ for an
initial state,  the $(m_1, m_2)_f$ for the final state generated by the
action of $\co$ is uniquely defined and therefore, the bivariate moments
defined by Eq. \eqref{eq.bbd4} are proper bivariate moments and they are
defined by the initial $(m_1, m_2)_i$. For completeness, let us mention
that  given the marginal centroids $(\epsilon_i , \epsilon_f)$, widths 
$(\sigma_i, \sigma_f)$ and the bivariate correlation coefficient 
$\zeta_{biv}$, the normalized bivariate Gaussian is defined by,
\be
\barr{l}
\rho_{\mbox{biv}-\cg;\co}(E_i , E_f) = \rho_{\mbox{biv}-\cg;\co}
(E_i,E_f;\epsilon_i,\epsilon_f,
\sigma_i,\sigma_f,\zeta_{biv}) 
\\ \\
= \dis\frac{1}{2\pi \sigma_i \sigma_f \sqrt{(1-\zeta_{biv}^2)}} 
\\ \\
\times \exp-\dis\frac{1}{2(1-\zeta_{biv}^2)}
\l\{\l(\frac{E_i - \epsilon_i}{\sigma_i}\r)^2 
- 2\zeta_{biv} \l(\frac{E_i-\epsilon_i}{\sigma_i}
\r)\l(\frac{E_f-\epsilon_f}{\sigma_f}\r) + \l(\frac{E_f-\epsilon_f}{
\sigma_f}\r)^2\;\r\}\;.
\earr \label{eq.bivg}
\ee

\subsection{Formulas for the bivariate moments}
\label{c7s3s2}

Using binary correlation approximation, we derive formulas for  the first
four moments $\wtM_{PQ}((m_1$ $, m_2)_i)$, $P+Q \leq 4$ of  $I_\co^{(m_1,
m_2)_i, (m_1, m_2)_f}(E_i,E_f)$ for any $k_\co$ by representing $H(k_H)$ and
$\co(k_\co)$ operators by independent EGOEs and assuming $H(k_H)$ is a
$k_H$-body operator preserving $(m_1,m_2)$'s. Note that the ensemble
averaged $k_H$-particle matrix elements of $H(k_H)$ are $v_H^2(i,j)$ with
$i+j=k_H$ [see Eq. (\ref{eq.b11})]  and similarly the ensemble average of 
$(v_\co^{\gamma \delta})^2$ is $v_\co^2$. From now on, we use $(m_1, m_2)_i
=  (m_1,m_2)$. Using Eq. (\ref{eq.9}) and  applying the basic rules given by
Eqs. (\ref{eq.b2}) and (\ref{eq.b3}), we have 
\be 
\barr{rcl}
\wtM_{00}(m_1,m_2) & = & \overline{\lan \co^\dg(k_\co) \co(k_\co) 
\ran^{m_1,m_2}} \\ \\
& = & \dis\sum_{\gamma, \delta} \;
\overline{\l\{v_\co^{\gamma\delta}\r\}^2} \; 
\lan \delta_2^\dg(k_\co) \gamma_1(k_\co) \gamma_1^\dg(k_\co) \delta_2(k_\co)
\ran^{m_1,m_2} \\ \\
& = & v_\co^2 \; \dis\binom{\wm_1}{k_\co} \; \dis\binom{m_2}{k_\co} \;.
\earr \label{eq.9aa}
\ee
Trivially, $\wtM_{10}(m_1,m_2)$ and $\wtM_{01}(m_1,m_2)$ will be zero as 
$H(k_H)$ is represented by EGOE$(k_H)$. Thus, $\wtM_{PQ}(m_1,m_2)$ are
central moments. Moreover, by definition, all the odd-order moments, i.e.,
$\wtM_{PQ}(m_1,m_2)$ with  $\mod(P+Q,2) \neq 0$, will be zero. Now, the
$\wtM_{11}$ is given by,
\be
\barr{rcl}
\wtM_{11}(m_1,m_2)  & = & \overline{\lan \co^\dagger(k_\co) H(k_H) 
\co(k_\co) H(k_H) \ran^{m_1,m_2}} \\ \\
& = & 
v_\co^2 \;\;
\dis\sum_{\alpha_1, \beta_1, \alpha_2, \beta_2, \gamma_1,
\delta_2; \; i+j = k_H} v_H^2(i,j) \;
\lan \gamma_1^\dagger(k_\co) \alpha_1(i) \beta_1^\dagger(i) \gamma_1(k_\co) 
\beta_1(i) \alpha_1^\dagger(i) \ran^{m_1} 
\earr \label{eq.9a}
\ee
\be
\barr{rcl}
& \times &
\lan \delta_2(k_\co) \alpha_2(j) \beta_2^\dagger(j) \delta_2^\dagger(k_\co) 
\beta_2(j) \alpha_2^\dagger(j) \ran^{m_2} \;. \nonumber
\earr \label{eq.9a-1}
\ee
Then, contracting over the $\gamma^\dg\gamma$ and $\delta\delta^\dg$ 
operators, respectively in
the first and second traces in Eq. (\ref{eq.9a}) using Eqs. (\ref{eq.b4}) 
and (\ref{eq.b5}) appropriately, we have
\be
\barr{rcl}
\wtM_{11}(m_1,m_2) & = & v_\co^2 \;\; \dis\sum_{i+j=k_H} v_H^2(i,j) \;
\dis\binom{\wm_1-i}{k_\co} \dis\binom{m_2-j}{k_\co} \\ \\
& \times & T(m_1,N_1,i) \; T(m_2,N_2,j) \;.
\earr \label{eq.9b}
\ee
Note that the formulas for the functions $T(\cdots)$'s appearing in  Eq.
(\ref{eq.9b})  are given by Eqs. (\ref{eq.b6a}), (\ref{eq.b6b}) and
(\ref{eq.5b}). Similarly, the functions $F(\cdots)$'s appearing ahead are
given by Eqs.  (\ref{eq.b8}) and (\ref{eq.3}). For the marginal variances,
we have
\be
\barr{rcl}
\wtM_{20}(m_1,m_2)  & = &  \overline{\lan \co^\dagger(k_\co) \co(k_\co)
H^2(k_H) \ran^{m_1,m_2}} \\ \\
& = & \wtM_{00}(m_1,m_2) \; \overline{\lan H^2(k_H) \ran^{m_1,m_2}} \;,
\\ \\
\wtM_{02}(m_1,m_2)  & = &  \overline{\lan \co^\dagger(k_\co) H^2(k_H) 
\co(k_\co) \ran^{m_1,m_2}} \\ \\
& = & \wtM_{00}(m_1,m_2) \; \overline{\lan H^2(k_H) \ran^{m_1+k_\co, 
m_2-k_\co}} 
\;.
\earr \label{eq.9ab}
\ee
In Eq. (\ref{eq.9ab}), the ensemble averages of $H^2(k_H)$ are given by Eq.
(\ref{eq.b12}). Now, the correlation coefficient $\zeta_{biv}$ is
\be
\zeta_{biv}(m_1,m_2) = \dis\frac{\wtM_{11}(m_1,m_2)}
{\dis\sqrt{\wtM_{20}(m_1,m_2) \; \wtM_{02}(m_1,m_2)}} \;.
\label{eq.11}
\ee
Clearly, $\zeta_{biv}$ will be independent of $v_\co^2$. 

Proceeding further, we derive formulas for the fourth order moments 
$\wtM_{PQ}$,  $P+Q = 4$. The results are as follows. Firstly, for $(PQ) =
(40)$ and $(04)$, we have 
\be
\barr{rcl}
\wtM_{40}(m_1,m_2) & = & \overline{\lan \co^\dagger(k_\co) \co(k_\co)
H^4(k_H) \ran^{m_1,m_2}} \\ \\
& = & \wtM_{00}(m_1,m_2) \; \overline{\lan H^4(k_H) \ran^{m_1,m_2}} 
\;, \\ \\
\wtM_{04}(m_1,m_2) & = & \overline{\lan \co^\dagger(k_\co) 
H^4(k_H) \co(k_\co) \ran^{m_1,m_2}} \\ \\
& = & \wtM_{00}(m_1,m_2) \; \overline{\lan H^4(k_H) 
\ran^{m_1+k_\co,m_2-k_\co}} \;.
\earr \label{eq.10}
\ee
In Eq. (\ref{eq.10}), the ensemble averages of $H^4(k_H)$ are 
given by Eq. (\ref{eq.b14}). For $(PQ) = (31)$, we have
\be
\barr{rcl}
\wtM_{31}(m_1,m_2) & = & \overline{\lan \co^\dagger(k_\co) H(k_H) 
\co(k_\co)  H^3(k_H) \ran^{m_1,m_2}} \\ \\
& = & \overline{\bcon{\co^\dagger(k_\co)}{H(k_H)}{\co(k_\co)}{H(k_H)} 
\bcon{\co^\dagger(k_\co)H(k_H)\co(k_\co)H(k_H)}{H(k_H)}{}{H(k_H)}
\lan \co^\dagger(k_\co) H(k_H) \co(k_\co) 
H(k_H) H(k_H) H(k_H) \ran^{m_1,m_2}} \\ \\
& + & \overline{\bcon{\co^\dagger(k_\co)}{H(k_H)}{\co(k_\co)H(k_H)}{H(k_H)} 
\bcon[2ex]{\co^\dagger(k_\co)H(k_H)\co(k_\co)}{H(k_H)}{H(k_H)}{H(k_H)}
\lan \co^\dagger(k_\co) H(k_H) \co(k_\co) 
H(k_H) H(k_H) H(k_H) \ran^{m_1,m_2}} \\ \\
& + & \overline{\bcon{\co^\dagger(k_\co)}{H(k_H)}{\co(k_\co)H(k_H)H(k_H)}
{H(k_H)} \bcon[2ex]{\co^\dagger(k_\co)H(k_H)\co(k_\co)}{H(k_H)}{}{H(k_H)}
\lan \co^\dagger(k_\co) H(k_H) \co(k_\co) 
H(k_H) H(k_H) H(k_H) \ran^{m_1,m_2}} \;.
\earr \label{eq.11a}
\ee
First and last terms on RHS of Eq. (\ref{eq.11a}) are simple as
$\bcon{}{H}{}{H} HH$ can be taken out of the average and then we are left
with a term similar to $\wtM_{11}(m_1,m_2)$. For the second term, the
$\co^\dg$ and $\co$ operators are contracted across $H$ operator using Eqs.
\eqref{eq.b4} and \eqref{eq.b5} and then one is left with an average of the
form $\lan HGHG \ran$. These will give the final formula,
\be 
\barr{l}
\wtM_{31}(m_1,m_2) = 2\; \overline{\lan H^2(k_H) \ran^{m_1,m_2}} \;
\wtM_{11}(m_1,m_2) \\ \\
+ \overline{\bcon{\co^\dagger(k_\co)}{H(k_H)}{\co(k_\co)H(k_H)}{H(k_H)} 
\bcon[2ex]{\co^\dagger(k_\co)H(k_H)\co(k_\co)}{H(k_H)}{H(k_H)}{H(k_H)}
\lan \co^\dagger(k_\co) H(k_H) \co(k_\co) 
H(k_H) H(k_H) H(k_H) \ran^{m_1,m_2}} \\ \\
= 2\; \overline{\lan H^2(k_H) \ran^{m_1,m_2}} \;
\wtM_{11}(m_1,m_2) 
+ v_\co^2 \; \dis\sum_{i+j=k_H, t+u=k_H} \; v_H^2(i,j) \; v_H^2(t,u) \;
\\ \\ \times \dis\binom{m_2-j}{k_\co}
\; \dis\binom{\wm_1-i}{k_\co} \; F(m_1,N_1,i,t) \; F(m_2,N_2,j,u) \;.
\earr \label{eq.12}
\ee
Similarly, we have
\be 
\barr{l}
\wtM_{13}(m_1,m_2)  = \overline{\lan \co^\dagger(k_\co) H^3(k_H) \co(k_\co) 
H(k_H) \ran^{m_1,m_2}} \\ \\
= \overline{\bcon{\co^\dagger(k_\co)}{H(k_H)}{}{H(k_H)}
\bcon{\co^\dagger(k_\co) H(k_H) H(k_H)}{H(k_H)}{\co(k_\co)}{H(k_H)}
\lan \co^\dagger(k_\co) H(k_H) H(k_H) H(k_H) \co(k_\co) 
H(k_H) \ran^{m_1,m_2}} \\ \\
+ \overline{\bcon{\co^\dagger(k_\co)}{H(k_H)}{H(k_H)}{H(k_H)}
\bcon[2ex]{\co^\dagger(k_\co) H(k_H)}{H(k_H)}{H(k_H) \co(k_\co)}{H(k_H)}
\lan \co^\dagger(k_\co) H(k_H) H(k_H) H(k_H) \co(k_\co) 
H(k_H) \ran^{m_1,m_2}} \\ \\
+ \overline{\bcon{\co^\dagger(k_\co)}{H(k_H)}{H(k_H)H(k_H)\co(k_\co)}
{H(k_H)}\bcon[2ex]{\co^\dagger(k_\co)H(k_H)}{H(k_H)}{}{H(k_H)}
\lan \co^\dagger(k_\co) H(k_H) H(k_H) H(k_H) \co(k_\co) 
H(k_H) \ran^{m_1,m_2}} 
\\ \\
=  2\; \overline{\lan H^2(k_H) \ran^{m_1+k_\co,m_2-k_\co}} \;
\wtM_{11}(m_1,m_2) \\ \\
+ v_\co^2 \; 
\dis\sum_{i+j=k_H, t+u=k_H}\; v_H^2(i,j) \; v_H^2(t,u) 
\; G(t,u) 
\\ \\
\times \dis\binom{\wm_1-k_\co-t+i}{i} \; \dis\binom{m_1+k_\co-t}{i}
\; \dis\binom{\wm_2-u+k_\co+j}{j} \; \dis\binom{m_2-k_\co-u}{j} \;;\\ \\
G(t,u) = \dis\binom{\wm_1-t}{k_\co} \dis\binom{m_2-u}{k_\co} T(m_1,N_1,t) 
\;  T(m_2,N_2,u)\;. 
\earr\label{eq.m13}
\ee
In Eq. \eqref{eq.m13}, the first and last terms can be evaluated by first
calculating the $H^2$ average over the intermediate states  $\l|
m_1+k_\co,m_2-k_\co \ran$ and then the remaining part is similar to 
$\wtM_{11}(m_1,m_2)$. Also, the second average is evaluated by first
contracting the two correlated $H$'s that are between $\co^\dg$ and $\co$
operators (see the contraction symbol for clarity) and then one is again
left with a term similar to $\wtM_{11}(m_1,m_2)$. Finally,
$\wtM_{22}(m_1,m_2)$ is given by,
\be
\barr{l}
\wtM_{22}(m_1,m_2) = \overline{\lan \co^\dagger(k_\co) H^2(k_H) \co(k_\co) 
H^2(k_H) \ran^{m_1,m_2}} \\ \\
= \overline{\bcon{\co^\dagger(k_\co)}{H(k_H)}{}{H(k_H)}
\bcon{\co^\dagger(k_\co) H(k_H) H(k_H) \co(k_\co)}{H(k_H)}{}{H(k_H)}
\lan \co^\dagger(k_\co) H(k_H) H(k_H) \co(k_\co) 
H(k_H) H(k_H) \ran^{m_1,m_2}} \\ \\
+ \overline{\bcon{\co^\dagger(k_\co)}{H(k_H)}{H(k_H) \co(k_\co)}{H(k_H)}
\bcon[2ex]{\co^\dagger(k_\co) H(k_H)}{H(k_H)}{\co(k_\co) H(k_H)}{H(k_H)}
\lan \co^\dagger(k_\co) H(k_H) H(k_H) \co(k_\co) 
H(k_H) H(k_H) \ran^{m_1,m_2}} \\ \\
+ \overline{\bcon{\co^\dagger(k_\co)}{H(k_H)}{H(k_H) \co(k_\co) H(k_H)}
{H(k_H)}\bcon[2ex]{\co^\dagger(k_\co) H(k_H)}{H(k_H)}{\co(k_\co)}{H(k_H)}
\lan \co^\dagger(k_\co) H(k_H) H(k_H) \co(k_\co) 
H(k_H) H(k_H) \ran^{m_1,m_2}} \nonumber
\earr \label{eq.m14-1}
\ee
\be
\barr{l}
=  \wtM_{00}(m_1,m_2) \; \overline{\lan H^2(k_H)
\ran^{m_1+k_\co,m_2-k_\co}} \;\; \overline{\lan H^2(k_H) \ran^{m_1,m_2}} 
\\ \\
+ v_\co^2 \; \dis\sum_{i+j=k_H, t+u=k_H}\; v_H^2(i,j) \; v_H^2(t,u) \;
\dis\binom{\wm_1-i-t}{k_\co} \; \dis\binom{m_2-u-j}{k_\co} 
\\ \\
\times \l[ F(m_1,N_1,i,t) \; F(m_2,N_2,j,u) \r. 
\\ \\ \l. + 
T(m_1,N_1,i) \; T(m_1,N_1,t)\; T(m_2,N_2,j) \; T(m_2,N_2,u) \r] \;.
\earr \label{eq.m14}
\ee
In Eq. \eqref{eq.m14}, the first term is evaluated by first calculating the
$H^2$ average (for the $H^2$ between $\co^\dg$ and $\co$ operators) over
the intermediate state $\l| m_1+k_\co,m_2-k_\co \ran$ and then one is left
with product of averages of $H^2$ and $\co^\dg\co$ operators. For the third
term, first the $\co^\dg$ and $\co$ operators are contracted across $H^2$
operator and then we are left with average of the form 
$\lan H^2 \ran \times \lan H^2 \ran$. Similarly, for the
second term, after contracting the $\co^\dg$ and $\co$ operators across
$H^2$ operator, we are left with an average of the form $\lan HGHG \ran$. 

\subsection{Numerical results for bivariate correlation coefficient and 
fourth order cumulants}
\label{c7s3s3}

Firstly, given the $\wtM_{PQ}(m_1,m_2)$, the normalized central moments 
$M_{PQ}$  are $M_{PQ}=\wtM_{PQ}/\wtM_{00}$. Then, the scaled moments 
$\widehat{M}_{PQ}$ are 
\be
\widehat{M}_{PQ} = \dis\frac{M_{PQ}(m_1,m_2)}
{\l[M_{20}(m_1,m_2)\r]^{P/2} \l[
M_{02}(m_1,m_2)\r]^{Q/2}}\;;\;\;\;\;
P+Q \geq 2\;.
\label{eq.20}
\ee
Now the fourth order cumulants are \cite{St-87},
\be
\barr{l}
k_{40}(m_1,m_2) = \widehat{M}_{40}(m_1,m_2) - 3\;, 
k_{04}(m_1,m_2) = \widehat{M}_{04}(m_1,m_2) - 3\;, \\
k_{31}(m_1,m_2) = \widehat{M}_{31}(m_1,m_2) - 3\; 
\widehat{M}_{11}(m_1,m_2)\;, \\
k_{13}(m_1,m_2) = \widehat{M}_{13}(m_1,m_2) - 3\; 
\widehat{M}_{11}(m_1,m_2)\;, \\
k_{22}(m_1,m_2) = \widehat{M}_{22}(m_1,m_2) - 2\; 
\widehat{M}_{11}^2(m_1,m_2) -1 \;. 
\earr \label{eq.21}
\ee

\begin{table}[htp] 
\setlength{\tabcolsep}{20pt}
\caption{Correlation coefficients $\zeta_{biv}(m_1,m_2)$ for some nuclei 
with $k_\co = 2$ as appropriate for $0\nu\beta^-\beta^-$ decay operator. 
Note that space \#1 is for protons and space \#2 for neutrons. 
The configuration spaces corresponding to $N_1$ or
$N_2 = 20$, 22, 30, 32, 44 and 58 are $r_3f$, $r_3g$, $r_4g$, $r_4h$, 
$r_5i$, and $r_6j$, respectively with $f$ = $^1f_{7/2}$, $g$ = $^1g_{9/2}$, 
$h$ = $^1h_{11/2}$, $i$ = $^1i_{13/2}$, $j$ = $^1j_{15/2}$, 
$r_3$ = $^1f_{5/2}$ $^2p_{3/2}$ $^2p_{1/2}$, $r_4$ =
$^1g_{7/2}$ $^2d_{5/2}$ $^2d_{3/2}$ $^3s_{1/2}$, 
$r_5$ = $^1h_{9/2}$ $^2f_{7/2}$ $^2f_{5/2}$ $^3p_{3/2}$
$^3p_{1/2}$ and $r_6$ = $^1i_{11/2}$ $^2g_{9/2}$ $^2g_{7/2}$ $^3d_{5/2}$ 
$^3d_{3/2}$ $^4s_{1/2}$. See text for details.}
\begin{center}
\begin{tabular}{cccccc}
\hline
Nuclei & $N_1$ & $m_1$ & $N_2$ & $m_2$ & $\zeta_{biv}(m_1,m_2)$ \\ 
\hline
$^{76}_{32}$Ge$_{44}$  & $22$ & $4 $ & $22$ & $16$ & $0.64$ \\ 
$^{82}_{34}$Se$_{48}$  & $22$ & $6 $ & $22$ & $20$ & $0.6$ \\ 
$^{100}_{42}$Mo$_{58}$ & $30$ & $2 $ & $32$ & $8 $ & $0.57$ \\
$^{128}_{52}$Te$_{76}$ & $32$ & $2 $ & $32$ & $26$ & $0.62$ \\
$^{130}_{52}$Te$_{78}$ & $32$ & $2 $ & $32$ & $28$ & $0.58$ \\
$^{150}_{60}$Nd$_{90}$ & $32$ & $10$ & $44$ & $8 $ & $0.72$ \\
$^{154}_{62}$Sm$_{92}$ & $32$ & $12$ & $44$ & $10$ & $0.76$ \\
$^{180}_{74}$W$_{106}$ & $32$ & $24$ & $44$ & $24$ & $0.77$ \\
$^{238}_{92}$U$_{146}$ & $44$ & $10$ & $58$ & $20$ & $0.83$ \\
\hline
\end{tabular}
\label{corr-coef}
\end{center}
\end{table}

\begin{table}[htp] 
\caption{Cumulants $k_{PQ}$, $P+Q =4$ for  some nuclei listed in Table
\ref{corr-coef}.  The  numbers in the brackets are for the strict dilute 
limit as explained in the text.  Just as in the construction of Table 
\ref{corr-coef}, we use $v_H^2(i,j)$  independent of $(i,j)$. See Table 
\ref{corr-coef} and text for details.}
\begin{center}
\begin{tabular}{cccccccccc}
\hline
Nuclei & $N_1$ & $m_1$ & $N_2$ & $m_2$ & $k_{40}$ & $k_{04}$ & $k_{13}$
& $k_{31}$ & $k_{22}$ \\ 
\hline
$^{100}_{42}$Mo$_{58}$ & $30$ & $2 $ & $32$ & $8 $ & $-0.45(-0.39)$ 
& $-0.42(-0.38)$ & $-0.24(-0.23)$ & $-0.26(-0.25)$ & $-0.20(-0.22)$ \\
$^{150}_{60}$Nd$_{90}$ & $32$ & $10$ & $44$ & $8 $ & $-0.27(-0.22)$ 
& $-0.29(-0.23)$ & $-0.22(-0.18)$ & $-0.20(-0.17)$ & $-0.19(-0.18)$ \\
$^{154}_{62}$Sm$_{92}$ & $32$ & $12$ & $44$ & $10$ & $-0.24(-0.18)$ 
& $-0.25(-0.18)$ & $-0.19(-0.15)$ & $-0.18(-0.15)$ & $-0.17(-0.15)$ \\
$^{180}_{74}$W$_{106}$ & $32$ & $24$ & $44$ & $24$ & $-0.19(-0.08)$ 
& $-0.20(-0.08)$ & $-0.17(-0.08)$ & $-0.15(-0.08)$ & $-0.15(-0.08)$ \\
$^{238}_{92}$U$_{146}$ & $44$ & $10$ & $58$ & $20$ & $-0.18(-0.13)$ 
& $-0.18(-0.13)$ & $-0.15(-0.11)$ & $-0.15(-0.11)$ & $-0.13(-0.11)$ \\
\hline
\end{tabular}
\label{cumu}
\end{center}
\end{table}

Assuming $v_H^2(i,j)$ defining $H(2)$ are independent of $(i,j)$ so that
$\zeta_{biv}$ is independent of $v_H^2$, we have calculated the value of
$\zeta_{biv}$ with $k_\co = 2$ for several $0\nu\beta^-\beta^-$ decay
candidate nuclei using Eq. (\ref{eq.11}) along with Eqs. (\ref{eq.9aa}),
(\ref{eq.9b}), (\ref{eq.9ab}) and (\ref{eq.b12}). For the function
$T(\cdots)$, we use Eq. (\ref{eq.b6a}). Note that $v_H^2(i,j)$ correspond to
the variance of two-particle matrix elements from the $p-p$ $(i=2,j=0)$, 
$n-n$
$(i=0,j=2)$ and $p-n$ $(i=1,j=1)$ interactions. Results are given in   Table
\ref{corr-coef}. It is seen that $\zeta_{biv} \sim 0.6-0.8$. It is important
to mention that $\zeta_{biv} = 0$ for GOE.  Therefore, the transition
strength density will be narrow in $(E_i,E_f)$ plane. In order to establish
the bivariate Gaussian form for the $0\nu\beta^-\beta^-$ decay transition
strength density, we have examined $k_{PQ}$, $P+Q =4$. For a good bivariate
Gaussian, $|k_{PQ}| \lazz 0.3$. Using Eqs. (\ref{eq.9aa}), (\ref{eq.9b}),
(\ref{eq.9ab}), (\ref{eq.10}), (\ref{eq.12})-(\ref{eq.21}) along with Eqs.
(\ref{eq.b12}) and (\ref{eq.b14}), we have calculated the cumulants
$k_{PQ}(m_1,m_2)$, $P+Q = 4$. These involve $T(\cdots)$ and $F(\cdots)$
functions. For set \#1 calculations in Table \ref{cumu}, we use Eq.
(\ref{eq.b6a}) for  $T(\cdots)$ and Eq. (\ref{eq.3}) for $F(\cdots)$. For
the set \#2 calculations, shown in `brackets' in Table \ref{cumu}, we use
Eq. (\ref{eq.b6b}) for  $T(\cdots)$, Eq. (\ref{eq.b8}) for $F(\cdots)$ and
replace everywhere $\binom{\wm_i+r}{s} \to \binom{N_i}{s}$  for any $(r,s)$
with $i=1,2$. Then we have the strict dilute limit. We show in Table
\ref{cumu}, bivariate cumulants for five heavy nuclei for both sets of
calculations and they clearly establish that bivariate Gaussian is a good
approximation (similar tests are made for $\beta$ decay operator in Appendix
A). We have also examined this analytically in the dilute limit
with $N_1,N_2 \to \infty$ and assuming $v_H^2(i,j)$ independent of $(i,j)$.
With these, we have expanded $k_{PQ}$ in powers of  $1/m_1$ and $1/m_2$
using Mathematica. It is seen that all the  $k_{PQ}$, $P+Q =4$ behave as,
\be
k_{PQ} = - \dis\frac{4}{m_1} + O\l( \dis\frac{1}{m_1^2} \r)
+ O\l( \dis\frac{m_2^2}{m_1^3} \r) + \ldots \;.
\label{eq.22}
\ee
Therefore, for $m_1 >> 1$ and $m_2 << m_1^{3/2}$, the strength density
approaches bivariate Gaussian form in general. It is important to recall
that the strong dependence on $m_1$ in Eq. (\ref{eq.22}) is due to the
nature of the operator $\co$ i.e., $\co(k_\co) \l| m_1, m_2 \ran = \l| m_1 +
k_\co, m_2 - k_\co \ran$. Thus, we conclude that bivariate Gaussian form is
a good approximation for $0\nu\beta^-\beta^-$ decay transition strength
densities. With this, one can apply the formulation given in Sec. 
\ref{sdm} with
the bivariate correlation coefficient $\zeta_{biv}$ given by Eqs.
(\ref{eq.11}),  (\ref{eq.9ab}) and (\ref{eq.9b}). The values given by the
two-orbit binary correlation theory  for $\zeta_{biv}$  can be  used as
starting values in practical calculations.

\begin{sidewaystable} 
\setlength{\tabcolsep}{7pt}
\caption{Correlation coefficients $\zeta_{biv}(m_1,m_2)$ and  cumulants
$k_{PQ}$, $P+Q =4$ for  some nuclei relevant for $\beta$ decay. 
The first four nuclei in the table are relevant for
$\beta^-$ transitions, next four nuclei are relevant for electron capture
and the last two nuclei are relevant for $\beta^+$ transitions.  The 
numbers in the brackets for $k_{PQ}$ are for the strict dilute limit as in
Table \ref{cumu}. We assume  $v_H^2(i,j)$ are independent of $(i,j)$ as used
in the calculations generating Tables \ref{corr-coef} and \ref{cumu}.
Here, $m_1=m_p,\;m_2=m_n$ for the first four nuclei and $m_1=m_n,\;m_2=m_p$
for the next six nuclei. See text for details.}
\begin{center}
\begin{tabular}{ccccccccccc}
\hline
Nuclei & $N_1$ & $m_1$ & $N_2$ & $m_2$ & $\zeta_{biv}(m_1,m_2)$ & $k_{40}$ 
& $k_{04}$ & $k_{13}$ & $k_{31}$ & $k_{22}$ \\ 
\hline
$^{62}_{27}$Co$_{35}$ & $20$ & $7 $ & $30$ & $15$ & $0.72$ & 
$-0.26(-0.18)$ & $-0.27(-0.18)$ & $-0.24(-0.16)$ & $-0.23(-0.16)$ & 
$-0.22(-0.16)$ \\
$^{64}_{27}$Co$_{37}$ & $20$ & $7 $ & $30$ & $17$ & $0.73$ & 
$-0.27(-0.16)$ & $-0.27(-0.16)$ & $-0.24(-0.15)$ & $-0.23(-0.15)$ & 
$-0.21(-0.15)$ \\
$^{62}_{26}$Fe$_{36}$ & $20$ & $6 $ & $30$ & $16$ & $0.72$ & 
$-0.28(-0.18)$ & $-0.28(-0.18)$ & $-0.24(-0.16)$ & $-0.24(-0.16)$ & 
$-0.22(-0.16)$ \\
$^{68}_{28}$Ni$_{40}$ & $20$ & $8 $ & $30$ & $20$ & $0.72$ & 
$-0.27(-0.14)$ & $-0.27(-0.14)$ & $-0.24(-0.13)$ & $-0.23(-0.13)$ & 
$-0.21(-0.13)$ \\
$^{65}_{32}$Ge$_{33}$ & $36$ & $5 $ & $36$ & $4 $ & $0.55$ & 
$-0.45(-0.41)$ & $-0.46(-0.42)$ & $-0.35(-0.33)$ & $-0.34(-0.32)$ & 
$-0.34(-0.34)$ \\
$^{69}_{34}$Se$_{35}$ & $36$ & $7 $ & $36$ & $6 $ & $0.66$ & 
$-0.36(-0.29)$ & $-0.34(-0.30)$ & $-0.28(-0.25)$ & $-0.28(-0.25)$ & 
$-0.27(-0.25)$ \\
$^{73}_{36}$Kr$_{37}$ & $36$ & $9 $ & $36$ & $8 $ & $0.72$ & 
$-0.28(-0.23)$ & $-0.28(-0.23)$ & $-0.24(-0.20)$ & $-0.24(-0.20)$ & 
$-0.23(-0.20)$ \\
$^{77}_{38}$Sr$_{39}$ & $36$ & $11$ & $36$ & $10$ & $0.76$ & 
$-0.24(-0.19)$ & $-0.24(-0.19)$ & $-0.21(-0.17)$ & $-0.21(-0.17)$ & 
$-0.20(-0.17)$ \\
$^{85}_{42}$Mo$_{43}$ & $36$ & $15$ & $36$ & $14$ & $0.79$ & 
$-0.20(-0.14)$ & $-0.21(-0.14)$ & $-0.19(-0.13)$ & $-0.18(-0.13)$ & 
$-0.17(-0.13)$ \\
$^{93}_{46}$Pd$_{47}$ & $36$ & $19$ & $36$ & $18$ & $0.80$ & 
$-0.19(-0.11)$ & $-0.19(-0.11)$ & $-0.18(-0.10)$ & $-0.17(-0.10)$ & 
$-0.16(-0.10)$ \\
\hline
\end{tabular}
\label{cumu-beta}
\end{center}
\end{sidewaystable}

\section{Conclusions}

To summarize, by extending the binary correlation approximation method for
spinless embedded $k$-body ensembles to ensembles with proton-neutron
degrees of freedom that involves traces involving product of powers of two
different operators over two-orbit configurations  (either the operators
preserve the two-orbit symmetry or change a two-orbit configuration to a
unique final configuration), we have established that the transition
strength density generated by the two-body part of the Hamiltonian is a
bivariate Gaussian for transition operators $\co(k_\co)$ that change $k_\co$
number of neutrons to $k_\co$ number of protons. Towards this end, we have
derived formulas for the fourth order cumulants of the transition strength
density and calculated their values for some realistic examples; they are
found to vary from $\sim -0.4$ to $-0.1$. It is important to mention that
the embedding algebra for the EGOEs used is $U(N) \supset U(N_p) \oplus
U(N_n)$ [$p$ denotes `protons' and $n$ denotes `neutrons'] with the
Hamiltonian preserving the symmetry and the transition operator breaking the
symmetry in a particular way. We have also derived a formula for the fourth
order trace defining the correlation coefficient of the bivariate transition
strength density for the transition operator relevant for
$0\nu\beta^-\beta^-$ decay. For nuclei from $^{76}$Ge to $^{238}$U, the
bivariate correlation coefficient is found to vary from $\sim 0.6 - 0.8$ 
and these values can be used as a starting point for calculating nuclear
transition matrix elements for NDBD using the spectral distribution method
outlined in Sec. III. In future, it  is important to test the approximations
leading to Eq. (\ref{eq.17a}) using shell model examples.  Although spectral
distribution method is expected to be valid in the chaotic domain of the
spectrum (usually away from the  ground state), it remains to be tested how
well the method applies to the calculation of NTME for NDBD. In the past,
the theory has been applied successfully for occupancies near the ground
state \cite{Po-91, Po-75, Ko-78, Sa-85}  and also it is shown that in the
level density analysis of heavy nuclei \cite{Fr-06a} that the theory extends
close to the ground state. In the near future, applications will be carried
out for NTME for some heavy nuclei ($^{100}$Mo, $^{154}$Sm, $^{150}$Nd,
$^{186}$W, $^{238}$U).

\acknowledgments

Thanks are due to S. Tomsovic, K. Kar, P.C. Srivastava, R. Sahu, R.U. Haq
and J. Farine for some useful discussions.

\renewcommand{\theequation}{A\arabic{equation}}
\setcounter{equation}{0}   

\section*{APPENDIX A} 

For completeness, we have also calculated the correlation coefficient and
fourth order moments for the transition operator relevant for $\beta$ decay
[$k_\co = 1$ in Eq. (\ref{eq.9})]. Results are given in Table
\ref{cumu-beta}. For the first four nuclei (they are $\beta^-$ candidates)
in the table, N = Z = 20 is used as the core. Here, $N_1$ corresponds to 
$^1f_{7/2}$ $^1f_{5/2}$ $^2p_{3/2}$ $^2p_{1/2}$ and $N_2$ corresponds to 
$^1f_{7/2}$ $^1f_{5/2}$ $^2p_{3/2}$ $^2p_{1/2}$ $^1g_{9/2}$.  Similarly, for
the remaining six nuclei (they are electron capture or $\beta^+$
candidates), N = Z = 28 and $N_1$ and $N_2$ correspond to $^1f_{5/2}$
$^2p_{3/2}$ $^2p_{1/2}$ $^1g_{9/2}$ $^1g_{7/2}$ $^2d_{5/2}$.  The fourth
order cumulants values presented in Table \ref{cumu-beta} confirm that the
bivariate Gaussian form is a good approximation for $\beta$ decay transition
strength densities. Results in Table \ref{cumu-beta} justify the assumptions
made  in \cite{Fr-88b,Ma-07} where spectral distribution method is applied,
with the correlation coefficients in the correct range, to calculate the
$\beta$ decay rates for nuclei relevant for pre-supernovae evolution.

\ed